\documentclass[12pt,thmsa]{article}
\usepackage{amssymb}

\usepackage{sw20aip}

\input{tcilatex}
\begin{document}

\author{Efrain J. Ferrer and Vivian de la Incera \\
\textit{Physics Department, State University of New York, Fredonia, NY 14063}}
\title{Ward-Takahashi Identity with External Field in Ladder QED$\thanks{%
Supported by NSF grant PHY-9722059}$}
\date{SUNY-FRE-98-03}
\maketitle

\begin{abstract}
We derive the Ward-Takahashi identity obeyed by the
fermion-antifermion-gauge boson vertex in ladder QED in the presence of a
constant magnetic field. The general structure in momentum space of the
fermion mass operator with external electromagnetic field is discussed.
Using it we find the solutions of the ladder WT identity with magnetic
field. The consistency of our results with the solutions of the
corresponding Schwinger-Dyson equation ensures the gauge invariance of the
magnetic field induced chiral symmetry breaking recently found in ladder QED.
\end{abstract}

\section{Introduction}

Recently, the problem of finding a mechanism to generate galactic magnetic
fields has undergone a boost of attention\cite{1}$^{,}$\cite{2}$^{,}$\cite{3}
. The interest has been motivated by the experimental observation in our
own, as well as in other galaxies, of magnetic fields of about $10^{-6}$G on
scales of the order of 100 kpc. The connection between the direction of the
observed galactic magnetic fields and the rotation axis of the galaxy
suggests that they have been produced by a dynamo effect, in which a small
seed magnetic field is amplified by the turbulent motion of matter during
galaxy formation\cite{4} . This fact leads to the problem of the generation
of seed primordial fields, which has been addressed in many\cite{1}$^{-}$%
\cite{3} papers, where a large number of different generating mechanisms has
been proposed.

Parallel to these investigations, and perhaps partially motivated by them,
several fundamental studies on the effects produced by magnetic fields in
quantum field theories\cite{3-1}$^{-}$\cite{11-2} have been recently carried
out. We would like to refer to a particularly interesting effect of a
magnetic background, known as the catalysis of chiral symmetry breaking due
to a magnetic (chromomagnetic, hypermagnetic) field. The essence of this
effect lies in the dimensional reduction in the dynamics of fermion pairing
in the presence of a magnetic field\cite{5} . Due to such a dimensional
reduction, the magnetic field catalyses the generation of a fermion
condensate, and consequently, of a dynamical fermion mass, even in the
weakest attractive interaction between fermions.

An important aspect of the magnetic field induced chiral symmetry breaking
(MCSB) is related to its possible cosmological consequences. It has been
speculated\cite{5} that the character of electroweak phase transition could
be affected by the MCSB. The existence of a broken chiral symmetry phase due
to a magnetic field in the early universe would require the presence of very
large primordial magnetic fields at those stages.

As mentioned above, the existence of primordial magnetic fields in the early
universe cannot be disregarded, but on the contrary, it seems to be needed
to explain the observed large-scale galactic magnetic fields. Several
primordial field generating mechanisms predict fields as large as 10$^{24}G$
during the electroweak phase transitions. Moreover, Ambj\o rn and Olesen\cite
{11-3} have claimed that seed fields even larger, $\sim 10^{33}G,$ would be
necessary at the electroweak phase transition to explain the observed
galactic fields. Notwithstanding the possible existence of such large seed
fields, it has been argued\cite{7}$^{,}$\cite{10} that they are too small to
make the MCSB an important effect during the electroweak phase transition.
This conclusion was drawn from the study of MCSB in QED at finite
temperature. Nevertheless, in recent calculations of the catalysis of chiral
symmetry breaking in an Abelian gauge model with Yukawa terms (which is a
model with many of the attributes of the fermion sector of the electroweak
theory), it has been found\cite{11-4} that the critical magnetic field,
required for this effect to be important at the electroweak scale, can be
substantially decreased by a Yukawa coupling of the order of the one
corresponding to the top quark. From our point of view, these results
indicate that the problem of the influence of the MCSB in the electroweak
transition is still open and deserve more investigation.

The catalysis of chiral symmetry breaking by a magnetic field has been found
in a variety of theories as Nambu-Jona-Lasinio\cite{5}$^{,}$\cite{9-1}$^{,}$%
\cite{11-1} , QED\cite{5}$^{-}$\cite{9-2} , and QCD\cite{9}$^{,}$\cite{11-2}
. These studies have been done using a Bethe-Salpeter\cite{5} (BS), or a
Schwinger-Dyson (SD) approach\cite{6}$^{-}$\cite{8} , at zero\cite{5}$^{,}$%
\cite{6} and at finite temperature\cite{7}$^{,}$\cite{10} , as well as in
QED at finite density\cite{8} (i.e. in the presence of chemical potential).
In all cases the dynamical mass was the result of a nonperturbative
calculation defined by the ladder and/or improved ladder approximations.

We must recall however that the so called ladder or rainbow approximation is
not gauge invariant\cite{12} . One may wonder then, whether the magnetic
field induced chiral mass is for real or just an artifact of the used
approximation.

Indeed this question is not new. In the absence of any external field, a
similar problem\cite{13} is present in the strong coupling regime of QED,
where the critical coupling for chiral symmetry breaking, found by solving
the SD (or the Bethe-Salpeter) equation for the fermion propagator in the
ladder approximation, is gauge dependent\cite{12} . The solution to this
problem was obtained by investigating the Ward-Takahashi (WT) identities in
the ladder approximation. As known, if the WT identities are satisfied by
the solution of the SD equations in some approximation in a certain gauge,
one can use the gauge transformation law for the Green's functions\cite{14}
to rewrite the SD equations in an arbitrary gauge. The transformation law
guarantees that the WT identities are satisfied by the solutions of the SD
equations in all other gauges, although the approximation on which the SD
equations are solved may change. As it turns out, the gauge invariance of
the critical coupling found in ladder QED\cite{12} is ensured by the
consistency of the WT identity with the SD solution, taken in the same
ladder approximation in the Landau gauge.

When a magnetic field is present, one can follow a similar strategy to prove
the gauge invariance of the magnetic field induced chiral mass obtained in
ladder QED. This program is far from trivial though. On one hand, the
fermion Green function cannot be diagonalized in momentum coordinates by a
Fourier transform. To diagonalize it, one must work in the representation of
the eigenfunctions of the fermion mass operator\cite{18}$^{,}$\cite{16} . On
the other hand, the magnetic field provides the theory with an extra tensor $%
F_{\mu \nu }$, making the structure of the fermion mass operator much richer%
\cite{18}$^{,}$\cite{19} . The new structural terms modify the WT identities
of the theory, which now involve a larger number of scalar coefficients (the
scalar coefficients of the different independent structures of the mass
operator).

The main goal of the present paper is to establish the gauge invariance of
the magnetic field induced chiral mass found in ladder QED, taking into
account the above mentioned features of the theory in the presence of a
magnetic field. With this aim, we study the ladder WT identity for the
fermion-antifermion-gauge boson vertex in the presence of a magnetic field.
Recalling that the consistency of the WT and SD solutions should be
established within the same approximation, we solve the ladder WT identity
in the lowest Landau level (LLL). As shown below, the solutions of the WT
and SD equations are indeed compatible in the Feynman gauge within this
approximation.

The plan of the paper is the following. In Section I, we derive the momentum
representation of the WT identity for ladder QED in a magnetic field. To
solve the identity, we need to use the fermion mass propagator, whose
structure in the p-space is discussed in Section II. The WT solution is
finally found in Section III for fermions in LLL. As we discuss there, it is
satisfied by the solution, obtained in refs.[7],[8], of the corresponding SD
equation within the same approximation. Hence, the gauge invariance of the
chiral mass is proved. In Section IV we give the conclusions.

\section{Ward-Takahashi Identity in Ladder QED with External Magnetic Field}

Let us derive the WT identity in momentum space for the
fermion-antifermion-gauge boson vertex in QED in the presence of a constant
magnetic field. In coordinate representation the WT identity looks the same
as in the case without magnetic field, 
\begin{equation}
\partial _{\mu }^{z}\frac{\delta ^{3}\Gamma (0)}{\delta \overline{\Psi }%
\left( x\right) \delta \Psi (y)\delta A_{\mu }\left( z\right) }=ie\delta
^{(4)}(y-z)\frac{\delta ^{2}\Gamma (0)}{\delta \overline{\Psi }\left(
x\right) \delta \Psi (y)}-ie\delta ^{(4)}\left( x-z\right) \frac{\delta
^{2}\Gamma (0)}{\delta \overline{\Psi }\left( x\right) \delta \Psi (y)}
\label{e1}
\end{equation}
where $\Gamma $ is the generating functional for proper vertices, and the
zero in $\Gamma $ means that all average fields, but the magnetic one, are
set to zero. The existence of the external magnetic field is manifested in
the modification of the fermion propagator as discussed below.

The solutions of (\ref{e1}) depend on the approximation used for the vertex
and the inverse propagator. As discussed in the introduction, we are
interested in the study of the identity (\ref{e1}) in the same approximation
used in refs.[7],[8] to solve the SD equation for the fermion propagator.
Therefore, we need to use the so called rainbow or ladder approximation, in
which the vertex is taken bare,

\begin{equation}
\frac{\delta ^{3}\Gamma (0)}{\delta \overline{\Psi }\left( x\right) \delta
\Psi (y)\delta A^{\mu }\left( z\right) }=-e\delta ^{(4)}\left( x-z\right)
\delta ^{(4)}(y-z)\gamma _{\mu }  \label{e2}
\end{equation}
and the fermion inverse propagator is taken full (denoted by a bar over $%
S^{-1}$),

\begin{equation}
\frac{\delta ^{2}\Gamma (0)}{\delta \overline{\Psi }\left( x\right) \delta
\Psi (y)}=\overline{S}^{-1}(x,y)  \label{e3}
\end{equation}

In this non-perturbative approximation the WT identity (\ref{e1}) becomes

\begin{equation}
-\partial _{z}^{\mu }[\delta ^{(4)}\left( x-z\right) \delta
^{(4)}(y-z)\gamma _{\mu }]=i\delta ^{(4)}(y-z)\overline{S}^{-1}(x,y)-i\delta
^{(4)}\left( x-z\right) \overline{S}^{-1}(x,y)  \label{e11}
\end{equation}

We are interested in the momentum representation of this identity. Usually,
the transformation to momentum representation can be done by Fourier
transforming eq.(\ref{e1}). In the presence of an external background field,
however, the Fourier transform of the fermion Green's functions is not
diagonal in the p-space. Fortunately, in the case of a constant uniform
magnetic field H, the fermion propagator, and hence its inverse, can be
p-diagonalized by working in the so called $E_{p}$-representation\cite{16} .

The $E_{p}$-representation, introduced by Ritus \cite{16} in his studies of
QED with external electromagnetic fields, is based on the eigenfunctions of
the operator $(\gamma \cdot \Pi )^{2},$ 
\begin{equation}
(\gamma \cdot \Pi )^{2}\psi _{p}=\overline{p}^{2}\psi _{p}  \label{e5}
\end{equation}
The operator $(\gamma \cdot \Pi )^{2}$ commutes with the mass operator.
Hence, it is convenient to work with the mass operator expressed in the
representation defined by these eigenfunctions.

In the chiral representation in which $\gamma _{5}$ and $\tsum_{3}=i\gamma
^{1}\gamma ^{2}$ are diagonal with eigenvalues $\pm 1,$ the eigenfunctions $%
\psi _{p}$ are given by 
\begin{equation}
\psi _{p}=E_{p\sigma }(x)\omega _{\sigma
}=N(n)e^{i(p_{0}x^{0}+p_{2}x^{2}+p_{3}x^{3})}D_{n}(\rho )\omega _{\sigma }
\label{7a}
\end{equation}
where $\omega _{\sigma }$ are the proper bispinors of $\gamma _{5}$ and $%
\tsum_{3}$, $D_{n}(\rho )$ are the parabolic cylinder functions\cite{17}
with argument $\rho =\sqrt{2\left| eH\right| }(x_{1}-\frac{p_{2}}{eH})$ and
positive integer index

\begin{equation}
n=n(k,\sigma )\equiv k+\frac{eH\sigma }{2\left| eH\right| }-\frac{1}{2}%
,\quad n=0,1,2,...  \label{e7}
\end{equation}
$N(n)=(4\pi \left| eH\right| ^{\frac{1}{4}}/\sqrt{n!}$ is a normalization
factor. Here $p$ represents the set $(p_{0},p_{2,}p_{3},k),$ which
determines the eigenvalue $\overline{p}^{2}=-p_{0}^{2}+p_{3}^{2}+2\left|
eH\right| k$ in eq.(\ref{e5})$.$

The $E_{p}$-representation is obtained forming the eigenfunction-matrices
(see refs.[7, 8, 21] for details)

\begin{equation}
E_{p}(x)=\sum\limits_{\sigma }E_{p\sigma }(x)\Delta (\sigma ),  \label{e4}
\end{equation}

where 
\begin{equation}
\Delta (\sigma )=diag(\delta _{\sigma 1},\delta _{\sigma -1},\delta _{\sigma
1},\delta _{\sigma -1}),\qquad \sigma =\pm 1,
\end{equation}

It is easy to check that the $E_{p}$ functions are orthonormal

\begin{equation}
\int d^{4}x\overline{E}_{p^{\prime }}(x)E_{p}(x)=(2\pi )^{4}\widehat{\delta }%
^{(4)}(p-p^{\prime })\equiv (2\pi )^{4}\delta _{kk^{\prime }}\delta
(p_{0}-p_{0}^{\prime })\delta (p_{2}-p_{2}^{\prime })\delta
(p_{3}-p_{3}^{\prime })  \label{e8}
\end{equation}
as well as complete

\begin{equation}
\sum\limits_{k}\int \ dp_{0}dp_{2}dp_{3}E_{p}(x)\overline{E}_{p}(y)=(2\pi
)^{4}\delta ^{(4)}(x-y)  \label{e9}
\end{equation}
Here we used $\overline{E}_{p}(x)=\gamma ^{0}E_{p}^{\dagger }\gamma ^{0}.$

They also satisfy two important relations

\begin{equation}
\gamma \cdot \Pi E_{p}(x)=E_{p}(x)\gamma \cdot \overline{p}  \label{e9a}
\end{equation}

\begin{equation}
\int d^{4}x^{\prime }M(x,x^{\prime })E_{p}(x)=E_{p}(x)\widetilde{\tsum }_{A}(%
\overline{p})  \label{9b}
\end{equation}
$M(x,x^{\prime })$ is the matrix elements of the mass operator in coordinate
representation.\ Equation (\ref{9b}) is then the eigenvalue equation for the
mass operator, with $\widetilde{\sum }_{A}(\overline{p})$ being the
eigenvalue matrix of the mass operator in momentum space. Notice that the
eigenvalue matrix $\widetilde{\sum }_{A}(\overline{p})$ is evaluated in $%
\overline{p}=(p_{0},0,-sgn(eH)\sqrt{2\left| eH\right| k_{p}},p_{3})$.

Let us transform eq.(\ref{e11}) to momentum space. First, note that in eq.(%
\ref{e11}) the fermion inverse propagators do not depend on $z.$ Therefore,
the transformation to momentum coordinates for the variable $z$ can be done
with the usual Fourier transform. Multiplying (\ref{e11}) by $e^{iqz}$ and
integrating in $z$ one finds

\begin{equation}
\delta ^{(4)}(x-y)e^{iqx}q^{\mu }\gamma _{\mu }=e^{iqy}\overline{S}%
^{-1}(x,y)-e^{iqx}\overline{S}^{-1}(x,y)  \label{e10}
\end{equation}
If one now multiplies eq.(\ref{e10}) by the left by $\overline{E}_{p}(x),$
integrates in $x$, and then multiplies it by the right by $E_{r}(y)$, and
integrates in $y$, it takes the form

\begin{equation}
\int d^{4}xe^{iqx}\overline{E}_{p}(x)q^{\mu }\gamma _{\mu }E_{r}(x)=\int
d^{4}xd^{4}y(e^{iqy}-e^{iqx})\overline{E}_{p}(x)\overline{S}%
^{-1}(x,y)E_{r}(x)  \label{e12}
\end{equation}

Using the properties of the $E_{p}$ functions, the integral in the l-h-s of
the identity (\ref{e12}) can be done\cite{6}$^{,}$\cite{7} to obtain

\[
\int d^{4}xe^{iqx}\overline{E}_{p}(x)q^{\mu }\gamma _{\mu }E_{r}(x)=(2\pi
)^{4}\delta ^{(3)}(r+q-p)e^{iq_{1}(r_{2}+p_{2})/2eH}e^{-\widehat{q}_{\perp
}^{2}/2}x\times 
\]
\begin{equation}
\times \sum_{\sigma _{p}\sigma _{r}}\frac{1}{\sqrt{n_{p}!n_{r}!}}%
e^{isgn(eH)(n_{p}-n_{r})\varphi }J_{n_{p}n_{r}}(\widehat{q}_{\perp })\Delta
_{p}q^{\mu }\gamma _{\mu }\Delta _{r}  \label{e13}
\end{equation}
In (\ref{e13}) we used the following notation

\begin{equation}
J_{n_{p}n_{r}}(\widehat{q}_{\perp })\equiv \sum\limits_{m=0}^{\min
(n_{p},n_{r})}\frac{n_{p}!n_{r}!}{m!(n_{p}-m)!(n_{r}-m)!}[isgn(eH)\widehat{q}%
_{\perp }]^{n_{p}+n_{r}-2m}  \label{e14}
\end{equation}

\[
\delta ^{(3)}(r+q-p)\equiv \delta (r_{0}+q_{0}-p_{0})\delta
(r_{2}+q_{2}-p_{2})\delta (r_{3}+q_{3}-p_{3}) 
\]
\[
\Delta _{p}\equiv \Delta (\sigma _{p}),\quad n_{p}=n(k_{p},\sigma
_{p}),\qquad \Delta _{r}=\Delta (\sigma _{r}),\quad n_{r}=n(k_{r},\sigma
_{r}), 
\]
along with the dimensionless variables

\begin{equation}
\widehat{q}_{\mu }\equiv \frac{q_{\mu }\sqrt{2\left| eH\right| }}{2eH}\
,\qquad \mu =0,1,2,3  \label{e15}
\end{equation}
and their corresponding polar coordinates

\begin{equation}
\widehat{q}_{\perp }\equiv \sqrt{\widehat{q}_{1}^{2}+\widehat{q}_{2}^{2}}%
,\quad \varphi \equiv \arctan (\widehat{q}_{2}/\widehat{q}_{1}),
\label{e-15a}
\end{equation}

Introducing now the function

\begin{equation}
\Phi _{n_{p}n_{r}}(r,q,p)\equiv (2\pi )^{4}\delta
^{(3)}(r+q-p)e^{iq_{1}(r_{2}+p_{2})/2eH}e^{-\widehat{q}_{\perp }^{2}/2}\frac{%
1}{\sqrt{n_{p}!n_{r}!}}e^{isgn(eH)(n_{p}-n_{r})\varphi }J_{n_{p}n_{r}}(%
\widehat{q}_{\perp })  \label{e16}
\end{equation}
it is possible to express equation (\ref{e13}) in a more compact way,

\begin{equation}
\int d^{4}xe^{iqx}\overline{E}_{p}(x)q^{\mu }\gamma _{\mu
}E_{r}(x)=\sum_{\sigma _{p}\sigma _{r}}\Phi _{n_{p}n_{r}}(r,q,p)\Delta
_{p}\gamma _{\mu }q^{\mu }\Delta _{r}  \label{e17}
\end{equation}

One can work in a similar way with the right-hand side of (\ref{e12}) in
order to express it also in terms of the $\Phi $ functions (\ref{e16}). With
that aim, let us recall that the $E_{p}$ -transform\cite{16} of the inverse
fermion propagator satisfies

\begin{equation}
\overline{S}^{-1}(x,y)=\sum\limits_{k_{p}}\int \
dp_{0}dp_{2}dp_{3}E_{p}(x)[\gamma \cdot \overline{p}+\widetilde{\tsum }_{A}(%
\overline{p})]\overline{E}_{p}(y)  \label{e18}
\end{equation}
where $\widetilde{\tsum }_{A}(\overline{p})$ is the eigenvalue matrix of the
mass operator introduced in (\ref{9b}).

Substituting (\ref{e17})and (\ref{e18}) back in (\ref{e12}), we obtain

\[
\sum_{\sigma _{p}\sigma _{r}}\Phi _{n_{p}n_{r}}(r,q,p)\Delta _{p}\gamma
_{\mu }q^{\mu }\Delta _{r}=\int d^{4}xd^{4}y(e^{iqy}-e^{iqx})\overline{E}%
_{p}(x)\times 
\]
\begin{equation}
\times \{\sum\limits_{k^{\prime }}\int \ dp_{0}^{\prime }dp_{2}^{\prime
}dp_{3}^{\prime }E_{p^{\prime }}(x)[\gamma \cdot \overline{p^{\prime }}+%
\widetilde{\tsum }_{A}(\overline{p}^{\prime })]\overline{E}_{p^{\prime
}}(y)\}E_{r}(y)  \label{e19}
\end{equation}

There are two different types of integrals in space coordinates in the right
hand side of (\ref{e19}). One involves only the product of two $E_{p}$
functions, so it is solved using the orthonormality relation (\ref{e8}). The
other integral is of the form

\[
\int d^{4}xe^{iqx}\overline{E}_{p}(x)E_{p^{\prime }}(x)=(2\pi )^{4}\delta
^{(3)}(p^{\prime }+q-p)e^{iq_{1}(p_{2}^{\prime }+p_{2})/2eH}e^{-\widehat{q}%
_{\perp }^{2}/2}\times 
\]
\begin{equation}
\times \sum_{\sigma _{p}\sigma _{p^{\prime }}}\frac{1}{\sqrt{%
n_{p}!n_{p^{\prime }}!}}e^{isgn(eH)(n_{p}-n_{p^{\prime }})\varphi
}J_{n_{p}n_{p^{\prime }}}(\widehat{q}_{\perp })\Delta _{p}\Delta _{p^{\prime
}}=\sum_{\sigma _{p}\sigma _{p^{\prime }}}\Phi _{n_{p^{\prime }}n_{p^{\prime
}}}(p^{\prime },q,p)\Delta _{p}\Delta _{p^{\prime }}  \label{e19a}
\end{equation}

Integrating in $x$ and $y$ in (\ref{e19}) and using eqs. (\ref{e8}) and (\ref
{e19a}), we obtain 
\[
\sum_{\sigma _{p}\sigma _{r}}\Phi _{n_{p}n_{r}}(r,q,p)\Delta _{p}\gamma
_{\mu }q^{\mu }\Delta _{r}= 
\]
\[
=\sum\limits_{k^{\prime }}\int \ dp_{0}^{\prime }dp_{2}^{\prime
}dp_{3}^{\prime }\{(2\pi )^{4}\widehat{\delta }^{(4)}(p-p^{\prime })[\gamma
\cdot \overline{p^{\prime }}+\widetilde{\tsum }_{A}(\overline{p}^{\prime
})]\sum_{\sigma _{p^{\prime }}\sigma _{r}}\Phi _{n_{p^{\prime \prime
}}n_{r}}(r,q,p^{\prime })\Delta _{p^{\prime }}\Delta _{r}\}- 
\]

\begin{equation}
-\sum\limits_{k^{\prime }}\int \ dp_{0}^{\prime }dp_{2}^{\prime
}dp_{3}^{\prime }\{(2\pi )^{4}\widehat{\delta }^{(4)}(p^{\prime
}-r)\sum_{\sigma _{p}\sigma _{p^{\prime }}}\Phi _{n_{p}n_{p^{\prime
}}}(p^{\prime },q,p)\Delta _{p}\Delta _{p^{\prime }}[\gamma \cdot \overline{%
p^{\prime }}+\widetilde{\tsum }_{A}(\overline{p}^{\prime })]\}  \label{e20}
\end{equation}

Thanks to the delta functions, the evaluation of the integrals in $%
p_{i}^{\prime }$ and the sum in $k^{\prime }$ in the identity (\ref{e20}) is
straightforward. Thus,

\[
\sum_{\sigma _{p}\sigma _{r}}\Phi _{n_{p}n_{r}}(r,q,p)\Delta _{p}\gamma
_{\mu }q^{\mu }\Delta _{r}= 
\]
\begin{equation}
\sum_{\sigma _{p}\sigma _{r}}\Phi _{n_{p}n_{r}}(r,q,p)\delta _{\sigma
_{p}\sigma _{r}}\{[\gamma \cdot \overline{p}+\widetilde{\tsum }_{A}(%
\overline{p})]\Delta _{p}-\Delta _{p}[\gamma \cdot \overline{r}+\widetilde{%
\tsum }_{A}(\overline{r})]\}  \label{e21}
\end{equation}
where we used

\begin{equation}
\Delta (\sigma )\Delta (\sigma ^{\prime })=\delta _{\sigma \sigma ^{\prime
}}\Delta (\sigma )  \label{e22}
\end{equation}

The matrices $\Delta (\sigma )$ satisfy the commutation relations

\begin{equation}
\Delta \gamma _{_{\parallel }}^{\mu }=\gamma _{_{\parallel }}^{\mu }\Delta
,\qquad \qquad \qquad with\ \text{\quad }\gamma _{_{\parallel }}^{\mu
}=(\gamma ^{0},0,0,\gamma ^{3})  \label{e23}
\end{equation}

\begin{equation}
\Delta \gamma _{_{\perp }}^{\mu }=\gamma _{_{\perp }}^{\mu }(1-\Delta
),\qquad \qquad with\ \text{\quad }\gamma _{_{\perp }}^{\mu }=(0,\gamma
^{1},\gamma ^{2},0)  \label{e24}
\end{equation}

Therefore, the matrix structure in the l-h s of (\ref{e21}) can be written as

\begin{equation}
\Delta _{p}\gamma _{\mu }q^{\mu }\Delta _{r}=(\gamma _{_{\perp }}\cdot
q_{_{\perp }})\Delta _{r}(1-\delta _{\sigma _{p}\sigma _{r}})+(\gamma
_{_{_{\parallel }}}\cdot q_{_{_{\parallel }}})\Delta _{r}\delta _{\sigma
_{p}\sigma _{r}}  \label{e24a}
\end{equation}

Using the above equations, and recalling that the function $\Phi
_{n_{p}n_{r}}$contains a tridimensional delta function $\delta
^{(3)}(r+q-p), $ one can eliminate the term proportional to $\gamma
_{_{\parallel }}\cdot q_{_{\parallel }}$ with a similar term proportional to 
$\gamma _{_{\parallel }}$ $\cdot (p_{_{\parallel }}-r_{_{\parallel }})$ in
the r-h-s of (\ref{e21})$.$ Then, the final expression in momentum space of
the WT identity in ladder QED with external magnetic field becomes

\[
\sum_{\sigma _{p}\sigma _{r}}\Phi _{n_{p}n_{r}}(r,q,p)\{(\gamma _{_{\perp
}}\cdot q_{_{\perp }})\Delta _{r}(1-\delta _{\sigma _{p}\sigma
_{r}})-[\gamma _{_{\perp }}\cdot (\overline{p}_{_{\perp }}+\overline{r}%
_{_{\perp }})\Delta _{p}-\gamma _{_{\perp }}\cdot \overline{r}_{_{\perp
}}]\delta _{\sigma _{p}\sigma _{r}}- 
\]

\begin{equation}
-[\widetilde{\tsum }_{A}(\overline{p})\Delta _{p}-\Delta _{r}\widetilde{%
\tsum }_{A}(\overline{r})]\delta _{\sigma _{p}\sigma _{r}}\}=0  \label{e25}
\end{equation}

It is evident that the equation (\ref{e25}) is showing already some
characteristic signs of the theory in the presence of a magnetic field. For
instance, one can notice a separation between parallel- and
perpendicular-to-the field variables. It contains also extra structures,
(like those formed by the products of $\gamma $ matrices times the matrices $%
\Delta _{r}),$which are not present in the case without external field.
Unless the structure of the mass operator eigenvalue matrix $\widetilde{\sum 
}_{A}$ is specified, we cannot find the solution of this WT identity. In the
next section, we discuss the invariant structure of $\widetilde{\sum }_{A}$
in the presence of a constant magnetic field.

\section{Mass Operator Structure}

The structure of the fermion mass operator can be found on very general
grounds. Since the mass operator must be a $\gamma $-matrix based scalar
function, it can be represented as a sum of Lorentz contractions between the
tensorial structures that can be formed using the available vectors and
tensors of the theory, and the elements of the Clifford algebra generated by
the Dirac matrices and their products. The possible contractions are
restricted however by the invariances of QED. That is, the mass operator
must be invariant under Lorentz, C, P, and T transformations.

In QED, because the only tensorial structure available is the vector
momentum $p_{\mu }$, the above requirements lead to the familiar structure

\begin{equation}
\widetilde{\tsum }(p)=Z\gamma \cdot p+mI,  \label{e26}
\end{equation}

Here $Z$ and $m$ are functions of the magnitude of $p$. Their values depend
on the approximation used to calculate $\widetilde{\tsum }$.

When a constant external electromagnetic field is present, in addition to
the momentum $p_{\mu }$, the theory contains the field strength $F_{\mu \nu
} $. Then, in the presence of an electromagnetic field, the general
structure of the mass operator becomes\cite{18}$^{,}$\cite{19}

\begin{equation}
\widetilde{\tsum }_{A}(p,F)=i\gamma \cdot V+mI+\sigma ^{\mu \nu }T_{\mu \nu
}+i\gamma _{_{5}}\gamma ^{\mu }A_{\mu }+R\gamma _{_{5}}F_{\mu \nu
}^{*}F^{^{_{\mu \nu }}}  \label{e27}
\end{equation}
where, besides the new term with the pseudoscalar $F_{\mu \nu
}^{*}F^{^{_{\mu \nu }}},$ we have now contractions with a vector $V_{\mu }$,
a pseudovector $A_{\mu }$, and a tensor $T_{\mu \nu }$, given respectively by

\begin{equation}
V_{\mu }=Zp_{\mu }+Z^{\prime }e^{2}F_{\mu \nu }F^{\nu \rho }p_{\rho }
\label{e28}
\end{equation}

\begin{equation}
A_{\mu }=AeF_{\mu \nu }^{*}p^{\nu }  \label{e29}
\end{equation}

\begin{equation}
T_{\mu \nu }=TeF_{\mu \nu }+T^{\prime }e(p_{\mu }F_{\nu \rho }-p_{\nu
}F_{\mu \rho })p_{\rho }  \label{e30}
\end{equation}

It can be checked that (\ref{e27}) satisfies all the required invariances of
the mass operator in QED. It should be pointed out, however, that the second
term in (\ref{e30}), was not considered in paper [20]. This term was
introduced by first time in ref.[22], since the structure it generates when
contracted with $\sigma ^{\mu \nu }$ satisfies all the invariance
requirements, and therefore, there is no reason to leave it out of the
general structure of $\widetilde{\tsum }_{A}$. Note that the structure $%
i\gamma \cdot (Fp)$ is absent from eq.(\ref{e27}), since it violates charge
conjugation

\begin{equation}
C\gamma ^{0}\widetilde{\tsum }_{A}^{*}(p,F)\gamma ^{0}C^{-1}=\widetilde{%
\tsum }_{A}(p,-F)  \label{e30a}
\end{equation}

In eqs.(\ref{e27})-(\ref{e30}) the coefficients $Z,Z^{\prime },A,T,T^{\prime
},m$ , and $R$ are functions of the magnitudes of the momentum and the
external field strength.

For a constant magnetic background field, the last term of eq.(\ref{e27})
vanishes. If we assume that the magnetic field is along the 3-axis, the only
non-zero component of $F_{\mu \nu }$ is $F_{12}=H$. Then the mass operator (%
\ref{e27}) can be written as

\begin{equation}
\widetilde{\tsum }_{A}(p,F)=iZ_{\shortparallel }\gamma \cdot
p_{\shortparallel }+iZ_{_{\perp }}\gamma \cdot p_{_{\perp }}+mI+2eHS\Sigma
_{3}+iAeH\Sigma _{3}\gamma \cdot p_{\shortparallel }  \label{e31}
\end{equation}
where we used the following definitions and equations

\begin{equation}
Z_{\shortparallel }\equiv Z,  \label{e31-a}
\end{equation}
\begin{equation}
Z_{_{\perp }}\equiv Z-Z^{\prime }e^{2}H^{2},  \label{e31b}
\end{equation}
\begin{equation}
S\equiv (T-T^{\prime }p_{_{\perp }}^{2}),  \label{e31c}
\end{equation}
\begin{equation}
\Sigma _{3}=\sigma ^{12}=i\gamma ^{1}\gamma ^{2},  \label{e31a}
\end{equation}
\begin{equation}
\gamma ^{5}\gamma ^{0}=\Sigma _{3}\gamma ^{3},  \label{e32}
\end{equation}
\begin{equation}
\gamma ^{5}\gamma ^{3}=\Sigma _{3}\gamma ^{0},  \label{e33}
\end{equation}

It is timely to recall that our goal in this section is to find the
structure of the eigenvalue matrix of the mass operator in momentum
coordinates $\widetilde{\sum }_{A}(\overline{p}),$ as defined in eq.(\ref{9b}%
); so we can use it later to solve the WT identity (\ref{e25}). From the
discussion leading to eq.(\ref{e31}), it is obvious that the structure of $%
\widetilde{\tsum }_{A}(\overline{p})$ will be given by (\ref{e31}) with $p$
substituted by $\overline{p}$%
\begin{eqnarray}
\widetilde{\tsum }_{A}(\overline{p}) &=&iZ_{\shortparallel }(\overline{p}%
,F)\gamma \cdot \overline{p}_{\shortparallel }+iZ_{_{\perp }}(\overline{p}%
,F)\gamma \cdot \overline{p}_{_{\perp }}+m(\overline{p},F)I+  \nonumber \\
&&+2eHS(\overline{p},F)\Sigma _{3}+iA(\overline{p},F)eH\Sigma _{3}\gamma
\cdot \overline{p}_{\shortparallel }  \label{e34}
\end{eqnarray}

Note the separation between parallel and perpendicular variables, typical of
the theory with external magnetic field. Due to this separation we have two $%
Z$ coefficients, $Z_{\shortparallel }$ and $Z_{_{\perp }},$ compared to only
one in the case with none magnetic field. Another physically expected new
term is the one related to the spin-field interaction (the fourth term in (%
\ref{e34})). The interpretation of the new term with coefficient $A$ is less
transparent. However, as we show below, the ladder WT identity requires the $%
A$ coefficient to be zero, although it allows a solution on which the
spin-field interaction term is present.

\section{Solutions of the Ladder WT Identity in External Magnetic Field}

To find the solution of the WT identity (\ref{e25}), we just need to
substitute the mass operator (\ref{e34}) in (\ref{e25}), and then find the
values of the coefficients that solve the WT identity. As discussed in the
introduction, we are interested in determining if the solutions of the
ladder SD equation for the fermion propagator\cite{6}$^{,}$\cite{7} satisfy
the WT identity when this one is taken in the same approximation.

With that aim, let us express the WT identity in a more convenient way,
before making explicit use of eq.(\ref{e34}). First, note that the function $%
\Phi _{n_{p}n_{r}}(r,q,p),$ which appears multiplying all the terms in the
WT identity (\ref{e25}), contains an exponential factor $e^{-\widehat{q}%
_{\perp }^{2}/2}$ (see the definition of $\Phi _{n_{p}n_{r}}$, eq. (\ref{e16}%
)). Notice also that the dimensionless variable $\widehat{q}_{\perp }^{2}$
is related to the perpendicular momentum of the photon and is not linked
(not even through the $\delta ^{(3)}(r+q-p)$ of the function $\Phi
_{n_{p}n_{r}}(r,q,p)$) to the arguments $\overline{p}=(p_{0},0,-sgn(eH)\sqrt{%
2\left| eH\right| k_{p}},p_{3})$ and $\overline{r}=(r_{0},0,-sgn(eH)\sqrt{%
2\left| eH\right| k_{r}},r_{3})$ of the mass operator in eq.(\ref{e25}). For
large $\widehat{q}_{\perp }$ the exponential function $e^{-\widehat{q}%
_{\perp }^{2}/2}$ ensures that the identity is always approximately
satisfied if the mass operator is finite in any region of the arguments $%
\overline{p}$ and $\overline{r}$. This last assumption, although rather
restrictive, is in agreement with the results of Hong, Kim and Sin\cite{9-1}
for the behavior of the mass operator in the infrared and ultraviolet
regions.

For small $\widehat{q}_{\perp },$ one can approximate\cite{6}$^{,}$\cite{7}
the function $J_{n_{p}n_{r}}(\widehat{q}_{\perp })$ appearing in the
definition of $\Phi _{n_{p}n_{r}},$ eq.(\ref{e16}), as

\begin{equation}
J_{n_{p}n_{r}}(\widehat{q}_{\perp })\cong n_{p}!\delta _{n_{p}n_{r}}
\label{e35}
\end{equation}
to obtain 
\begin{equation}
\Phi _{n_{p}n_{r}}(r,q,p)\cong (2\pi )^{4}\delta
^{(3)}(r+q-p)e^{iq_{1}(r_{2}+p_{2})/2eH}e^{-\widehat{q}_{\perp
}^{2}/2}\delta _{n_{p}n_{r}}  \label{e36}
\end{equation}

Substituting (\ref{e36}) in the WT identity (\ref{e25}), it can be written as

\[
\sum_{\sigma _{p}\sigma _{r}}\delta _{k_{p}+\frac{sgn(eH)}{2}\sigma
_{p},k_{r}+\frac{sgn(eH)}{2}\sigma _{r}}\{(\gamma _{_{\perp }}\cdot
q_{_{\perp }})\Delta _{r}= 
\]

\begin{eqnarray}
&=&\sum_{\sigma _{p}\sigma _{r}}\delta _{\sigma _{p}\sigma _{r}}\delta
_{k_{p}+\frac{sgn(eH)}{2}\sigma _{p},k_{r}+\frac{sgn(eH)}{2}\sigma
_{r}}\{(\gamma _{_{\perp }}\cdot q_{_{\perp }})\Delta _{r}+\gamma _{_{\perp
}}\cdot (\overline{p}_{_{\perp }}+\overline{r}_{_{\perp }})\Delta
_{p}-(\gamma _{_{\perp }}\cdot \overline{r}_{_{\perp }})+  \nonumber \\
&&+\widetilde{\tsum }_{A}(\overline{p})\Delta _{p}-\Delta _{r}\widetilde{%
\tsum }_{A}(\overline{r})\}  \label{e37}
\end{eqnarray}
where we cancelled out the common functions in both sides. If we sum in $%
\sigma _{r}$ in the r-h-s of (\ref{e37}), and rename $\sigma _{p}$ as $%
\sigma $, we find

\[
\sum_{\sigma _{p}\sigma _{r}}\delta _{k_{p}+\frac{sgn(eH)}{2}\sigma
_{p},k_{r}+\frac{sgn(eH)}{2}\sigma _{r}}(\gamma _{_{\perp }}\cdot q_{_{\perp
}})\Delta _{r}= 
\]
\begin{equation}
=\sum_{\sigma }\delta _{_{_{k_{p}k_{r}}}}\{(\gamma _{_{\perp }}\cdot
q_{_{\perp }})\Delta (\sigma )+\gamma _{_{\perp }}\cdot (\overline{p}%
_{_{\perp }}+\overline{r}_{_{\perp }})\Delta (\sigma )-(\gamma _{_{\perp
}}\cdot \overline{r}_{_{\perp }})+\widetilde{\tsum }_{A}(\overline{p})\Delta
(\sigma )-\Delta (\sigma )\widetilde{\tsum }_{A}(\overline{r})\}  \label{e38}
\end{equation}

Summing now in all $\sigma $ variables, and using the relation $\Delta
(1)+\Delta (-1)=1$, it is straightforwardly found that the WT identity
reduces to 
\[
(\gamma _{_{\perp }}\cdot q_{_{\perp }})[\delta _{k_{p},k_{r}+sgn(eH)}\Delta
(1)+\delta _{k_{p},k_{r}-sgn(eH)}\Delta (-1)]= 
\]
\[
=\delta _{_{_{k_{p}k_{r}}}}\{iZ_{\shortparallel }(\overline{p},F)\gamma
\cdot \overline{p}_{\shortparallel }-iZ_{\shortparallel }(\overline{r}%
,F)\gamma \cdot \overline{r}_{\shortparallel }+i[Z_{_{\perp }}(\overline{p}%
,F)-Z_{_{\perp }}(\overline{r},F)]\gamma \cdot \overline{p}_{_{\perp }}+[m(%
\overline{p},F)-m(\overline{r},F)]I+ 
\]
\begin{equation}
+2eH[S(\overline{p},F)-S(\overline{r},F)]\Sigma _{3}+iA(\overline{p}%
,F)eH\Sigma _{3}\gamma \cdot \overline{p}_{\shortparallel }-iA(\overline{r}%
,F)eH\Sigma _{3}\gamma \cdot \overline{r}_{\shortparallel }  \label{e39}
\end{equation}
In eq.(\ref{e39}) we already substituted the mass operators appearing in (%
\ref{e38}) by their explicit expressions according to eq.(\ref{e34}), and
used that $\overline{p}_{_{\perp }}=$ $\overline{r}_{_{\perp }}$ in all
terms multiplied by $\delta _{_{_{k_{p}k_{r}}}}$.

The origin of the term in the l-h-s of (\ref{e39}) can be traced to the $%
E_{p}$ transform of the vertex. We might think that because of this term,
there are no self-consistent solutions of the ladder WT identity.
Fortunately, we have not yet considered all the restrictions of the
approximation used in the calculation of the ladder SD solutions\cite{6}$%
^{,} $\cite{7} . To work in the same approximation, we need to constraint
the fermions to the LLL. This means that we must evaluate (\ref{e39}) in $%
k_{p}=k_{r}=0.$

Therefore, for fermions in the LLL, the WT identity is simply equivalent to

\[
iZ_{\shortparallel }(\overline{p},F)\gamma \cdot \overline{p}%
_{\shortparallel }+m(\overline{p},F)I+2eHS(\overline{p},F)\Sigma _{3}+iA(%
\overline{p},F)eH\Sigma _{3}\gamma \cdot \overline{p}_{\shortparallel }= 
\]
\begin{equation}
=iZ_{\shortparallel }(\overline{r},F)\gamma \cdot \overline{r}%
_{\shortparallel }+m(\overline{r},F)I+2eHS(\overline{r},F)\Sigma _{3}+iA(%
\overline{r},F)eH\Sigma _{3}\gamma \cdot \overline{r}_{\shortparallel }
\label{e40}
\end{equation}

Note that in this approximation $Z_{_{\perp }}$ disappears from the WT
identity. The identity (\ref{e40}) is satisfied if 
\begin{equation}
Z_{\shortparallel }(\overline{p},F)=Z_{\shortparallel }(\overline{r},F)=0
\label{41}
\end{equation}
\begin{equation}
m(\overline{p},F)=m(\overline{r},F)  \label{42}
\end{equation}
\begin{equation}
S(\overline{p},F)=S(\overline{r},F)  \label{43}
\end{equation}
\begin{equation}
A(\overline{p},F)=A(\overline{r},F)=0  \label{44}
\end{equation}

Comparing the solutions (\ref{41})-(\ref{44}) for the WT identity with the
solutions found by Leung, Ng and Arkley\cite{6} and by Lee, Leung and Ng\cite
{7} for the SD equation in the Feynman gauge using the same approximation,
we see that they are consistent with each other. The SD solutions of these
papers\cite{6}$^{,}$\cite{7} can be summarized by

\begin{equation}
Z=Z_{\shortparallel }=Z_{_{\perp }}=0  \label{45}
\end{equation}
\begin{equation}
m(\overline{p},F)=m(\overline{r},F)\approx m  \label{46}
\end{equation}
The other coefficients were assumed to be zero 
\begin{equation}
S(\overline{p},F)=S(\overline{r},F)=0  \label{47}
\end{equation}
\begin{equation}
A(\overline{p},F)=A(\overline{r},F)=0  \label{48}
\end{equation}
what, although not the most general solution, is however compatible with (%
\ref{43}) and (\ref{44}).

We conclude in this way that the chiral symmetry breaking, found in ladder
QED through the SD approach in the LLL limit in the Feynman gauge, is
consistent with the corresponding WT identity. Therefore, the chiral
symmetry breaking obtained by Miransky\cite{5} , et. al., and by Leung\cite
{6}$^{,}$\cite{7} , et. al., is not an artifact of the used approximation,
but a gauge independent result.

The magnetic field induced chiral symmetry breaking has been also found in
non-Abelian models. To prove the gauge invariance of the magnetic field
induced chiral mass in a non-Abelian case, one would need to do a similar
study on the consistency of the SD solution with the solution of the
correspondent WT identity. However, the WT identity in this case must be
substituted by the Slavnov-Taylor identity for the fermion-antifermion-gluon
vertex taken in the same nonperturbative approximation on which the SD
equation is solved. The main difficulty in this case is that the
Slavnov-Taylor identity has not been extended beyond a perturbative analysis.

\section{Conclusions}

In this paper we have studied the WT identity for the
fermion--antifermion-gauge boson vertex in massless QED in the presence of a
constant magnetic field. The fermions Green functions were considered exact
in the external field, while the quantum effects were calculated within the
ladder approximation, i.e., taking the bare vertex and the free photon
propagator, but the full electron propagator. To find the WT identity in the
p-space, we worked in the $E_{p}$-representation, where the fermion mass
operator is diagonal in momentum coordinates.

Our approach led us to a nonperturbative WT identity involving the electron
mass propagator in the presence of a magnetic field. To solve the identity
we used the structure of the mass propagator in the p-space in the presence
of a magnetic field. The mass operator in the momentum space was found by
considering all the independent $\gamma $-matrix based scalars that could be
formed out of $p$, $F_{\mu \nu },$ and the elements of the Dirac Clifford
algebra$,$ which satisfy the symmetries of the theory.

For lowest Landau level fermions, we found an explicit solution of the
ladder WT identity. The LLL case is specially important, since according to
recent results in the problem of MCSB\cite{5}$^{-}$\cite{7} , it is
precisely the LLL contribution what determines the appearance of a fermion
mass in ladder QED with constant magnetic field. Since the solution of the
SD equation that led to the existence of a chiral mass is in agreement with
our general solution for the WT identity within the same approximation, we
have effectively established that the dynamical chiral symmetry breaking due
to a magnetic field is a gauge invariant result.

\begin{description}
\item  
\begin{quote}
\textbf{Acknowledgments}
\end{quote}
\end{description}

It is a pleasure to thanks Dr. D. Caldi, Dr. P. McGraw and Dr. Y. J. Ng for
useful discussions.

\end{document}